                                 \documentstyle[12pt,aaspp4]{article}
                                \begin{document}

                                \title{
Metallicity Gradients in the Intracluster Gas of Abell 496
                                }

                                \author{
Renato A. Dupke$^{1,2}$ \& Raymond E. White III$^{1,3}$
                                }

				\affil{
$^1$Department of Physics \& Astronomy,
University of Alabama, Tuscaloosa, AL 35487-0324
                                }
				\affil{
$^2$Department of  Astronomy, University of Michigan,
   Ann Arbor, MI~48109-1090
				}
                                 \affil{
$^3$Laboratory for High Energy Astrophysics, Code 662, NASA/GSFC, 
Greenbelt, MD 20771
                                }
                                \begin{abstract}
Analysis of spatially resolved {\sl ASCA} spectra of the intracluster gas
in Abell 496 confirms there are mild metal abundance enhancements near the 
center, as previously found by White et al.\ (1994)
in a joint analysis of {\sl Ginga} LAC and {\sl Einstein} SSS spectra.
Simultaneous analysis of spectra from all {\sl ASCA} instruments (SIS + GIS) 
shows that the iron abundance is $0.36 \pm 0.03$ solar $3-12'$ from
the center of the cluster and rises $\sim$50\% to $0.53 \pm  0.04$ solar 
within the central $2'$.
The $F$-test shows that this abundance gradient is significant at 
the $>$99.99\% level.
Nickel and sulfur abundances are also centrally enhanced.
We use a variety of elemental abundance ratios to assess the relative
contribution of SN Ia and SN II to the metal enrichment of the intracluster gas.
We find spatial gradients in several abundance ratios, indicating that
the fraction of iron from SN Ia increases toward the cluster center,
with SN Ia accounting for $\sim$50\% of the iron mass $3-12'$ from
the center and $\sim$70\% within $2'$.
The increased proportion of SN Ia ejecta at the center is such that
the central iron abundance enhancement can be attributed wholly to SN Ia;
we find no significant gradient in SN II ejecta.
These spatial gradients in the proportion of SN Ia/II ejecta imply
that the dominant metal enrichment mechanism near the center 
is different than in the outer parts of the cluster.
We show that the central abundance enhancement is unlikely to be due to 
ram pressure stripping of gas from cluster galaxies, or to secularly 
accumulated stellar mass loss within the central cD.
We suggest that the additional SN Ia ejecta near the center is the 
vestige of a secondary SN Ia-driven wind from the cD 
(following a more energetic
protogalactic SN II-driven wind phase), which was partially 
smothered in the cD due to its location at the cluster center.

                                \end{abstract}
 
                                \clearpage
                                \section{
Introduction
                                }
 
While the metals observed in intracluster gas clearly originate from stars, it 
remains controversial how the metals got from stars into the intracluster gas.
The two global metal enrichment mechanisms considered to be most likely are
protogalactic winds from early-type galaxies (Larson \& Dinerstein 1975)
and ram pressure stripping of gas from galaxies (Gunn \& Gott 1972).
Early $Einstein$ FPCS spectroscopy (Canizares et al.\ 1982) 
and more recent {\sl ASCA} spectroscopy 
(Mushotzky \& Loewenstein 1997; Mushotzky et al.\ 1996) showed that global 
intracluster metal abundances are consistent with ejecta 
from Type II supernovae, which supports the protogalactic wind model.  
White (1991) showed that the specific energy of intracluster gas is greater 
than that of cluster galaxies, which also suggests that protogalactic winds 
injected significant amounts of energy and metals into intracluster gas.
However, theoretical uncertainties about the elemental yields from 
Type II supernovae make it difficult to determine confidently 
the relative proportion of iron from SN II and SN Ia in intracluster 
gas (Gibson, Loewenstein \& Mushotzky 1997). 
This uncertainty has allowed others to conclude that as much as $\sim$50\% of 
the iron in intracluster gas comes from SN Ia 
(Ishimaru \& Arimoto 1997; Nagataki \& Sato 1998).
The possible presence of such large quantities of iron from SN Ia is 
problematic. 
Is ram pressure stripping so effective that it contaminates the outer 
parts of clusters nearly as effectively as the central regions?
Or is it ejecta from a secondary SN Ia-driven wind phase in ellipticals? 
Clues about the dominant enrichment mechanism(s) may be found in 
the detailed spatial distribution of elements in intracluster gas.

An increasing number of galaxy clusters are being found with centrally
enhanced metal abundances in their intracluster gas.
$Ginga$ observations of the Virgo cluster showed that its iron abundance 
declines from $\sim$0.5 solar at the center to $\sim$0.1--0.2 solar 
3$^\circ$ away (Koyama, Takano \& Tawara 1991).
White et al.\ (1994) found central abundance enhancements in Abell 496 and 
Abell 2142 in joint analyses of $Ginga$ LAC and $Einstein$ SSS spectra.
{\sl ASCA} observations of the Centaurus cluster show that its iron abundance 
declines from $\sim$ solar at the center 
to $\sim$0.3 solar 15$^\prime$ away (Fukazawa et al.\ 1994). 
The Perseus cluster also has an abundance gradient near its center
(Ulmer et al.\ 1987; Ponman et al.\ 1990;
Kowalski et al.\ 1993; Arnaud, et al.\ 1994).
More recently, central abundance enhancements were found in
{\sl ASCA} data for Hydra A (Ikebe et al.\ 1997), AWM 7 
(Ezawa et al.\ 1997; Xu et al.\ 1997), Abell 2199 and Abell 3571 (Dupke 1998, Dupke \& White 2000), 
as well as in {\sl ROSAT} (Pislar et al.\ 1997) 
and {\sl BeppoSax} (Irwin \& Bregman 1999) data for Abell 85.
The presence of these central abundance enhancements is poorly correlated
with global cluster properties.

In this paper we analyze spatially resolved {\sl ASCA} spectra of
Abell 496 and confirm that it has centrally enhanced metal abundances.
Abell 496 is a Bautz-Morgan Type I cluster with an optical redshift of 
$z=0.0328$.
Adopting a Hubble constant of 50 km s$^{-1}$ Mpc$^{-1}$ and $q_0=0.5$, its
luminosity distance is $197\ h_{50}^{-1}$ Mpc and $1^\prime=57\ h_{50}^{-1}$ 
kpc.
The central cD (MCG -02-12-039) has a total $B$ magnitude of $B_T=13.42$ 
(Valentijn 1983) and an optical effective radius of 
$r_{\rm eff} \geq 49\ h_{50}^{-1}$ kpc (Schombert 1986).
Neither the projected galaxy distribution nor the galaxy velocity distribution 
in the cluster 
shows signs of significant substructure, so the cluster appears to be
dynamically relaxed (Bird 1993; Zabludoff, Huchra \& Geller 1990).
Heckman (1981) was the first to suggest that Abell 496 contained a cooling flow,
after he detected cool H$\alpha$-emitting gas in its central cD galaxy;  
subsequent optical observations found extended H$\alpha$ 
emission (Cowie et al.\ 1983). 
Nulsen et al.\ (1982) found a soft X-ray component in $Einstein$ SSS spectra 
of this cluster and estimated a cooling accretion rate of 
$\sim$200 $M_\odot$ yr$^{-1}$, which is
consistent with later analyses (Mushotzky  1984; Mushotzky
\& Szymkowiak 1988; Canizares, Markert \& Donahue 1988;  
Thomas, Fabian \& Nulsen 1987; White et al.\ 1994).
In the course of a joint analysis of $Einstein$ SSS and $Ginga$ LAC spectra,
White et al.\ (1994) found a central abundance enhancement in Abell 496;
the differing fields of view of these two instruments allowed a coarsely          
spatially-resolved analysis.
The {\sl ASCA} observations of Abell 496 have been previously analyzed by
Mushotzky (1995) and Mushotzky et al.\ (1996), who found no evidence of an 
abundance gradient within the central 1 Mpc; however, the first paper analyzed 
data from only one of the four {\sl ASCA} spectrometers, while the second 
considered data from all four {\sl ASCA} spectrometers, but only beyond 
3$^\prime$ from the center (in order to avoid the spectral influence of 
the central cooling flow). 
We include the central cooling flow region in our analysis
of data from all four {\sl ASCA} spectrometers.

                                \section{
Data Reduction \& Analysis
                                }

Abell 496 was observed for 40 ksec by {\sl ASCA} on 20-21 September 1993.
{\sl ASCA} carries four large-area X-ray telescopes, each with its own 
detector: two Gas Imaging Spectrometers (GIS) and two Solid-State Imaging 
Spectrometers (SIS).
Each GIS has a $50'$ diameter circular field of view and a usable
energy range of 0.7--12 keV; each SIS has a
$22'$ square field of view  and a usable energy range of 0.4--10 keV.
We selected data taken with high and medium bit rates, with cosmic ray 
rigidity values $\ge$ 6 GeV/c, with elevation angles from the
bright Earth of $\ge20^{\circ}$ and from the
Earth's limb of $\ge5^{\circ}$ (GIS) or $10^{\circ}$ (SIS); we also
excluded times when the satellite was affected by the South Atlantic Anomaly.
Rise time rejection of particle events was performed on GIS data, and
hot and flickering pixels were removed from SIS data.
The resulting effective exposure times for each instrument are shown in Table 1.
Since the cluster fills the spectrometers' fields of view, we 
estimated the background from blank sky files provided by the {\sl ASCA} Guest 
Observer Facility.

We used XSPEC v9.0 (Arnaud 1996) software to analyze the SIS and GIS spectra
separately and jointly.
We fit spectra using the {\tt mekal} and {\tt vmekal} thermal emission models,
which are based on the emissivity calculations of Mewe \& Kaastra
(cf. Mewe, Gronenschild \& van den Oord  1985; Mewe, Lemen \& van den Oord 1986; 
Kaastra  1992), with Fe L calculations by Liedahl, Osterheld \& Goldstein 
(1995).
Abundances are measured relative to the solar photospheric values of 
Anders \& Grevesse (1989), in which Fe/H=$4.68\times10^{-5}$ by number.
Galactic photoelectric absorption was incorporated using the {\tt wabs} 
model (Morrison \& McCammon  1983);  
the Galactic column of absorbing material
in this line of sight is $N_H=4.58\times10^{20}$ cm$^{-2}$
(Dickey \& Lockman 1990; {\sl HEASARC} {\tt NH} software).
Spectral channels were grouped to have at least 25 counts/channel. Energy 
ranges were restricted to 0.8--10 keV for the GIS and 0.4--10 keV for the SIS.
The minimum projected sizes for regions of spectral extraction
(circular radii or annular widths) were typically 3$^\prime$ for the 
GIS and 2$^\prime$ for the SIS.
To maximize the statistical significance of any gradients, we also
assessed larger regions in the outer parts.
A central region of GIS data with a projected radius of 2$^\prime$ was also 
analyzed for closer comparison with the SIS. 
The intracluster gas temperature in Abell 496 is cool enough ($\sim$4 keV)
that the energy dependence of {\sl ASCA}'s point spread function 
does not affect our results significantly.
Since our results for spectral fits to the individual instruments are consistent
with our results from the joint analysis of all instruments, we will
discuss only the joint analysis.

                                \section{
Results                                }
				\subsection{
Temperature and Abundance Profiles
                                }
 
We jointly fitted thermal models to spectra from all four {\sl ASCA} 
instruments.
The fitting normalizations for the data from each instrument were allowed to vary 
independently, in order to compensate for small calibration and spatial 
extraction 
differences between the detectors; the normalizations differ by $<10\%$ in 
practice.
Figure 1a shows the GIS data with the best fits in both the inner and outer
projected regions of the cluster, while Figure 1b shows the same fits to the 
SIS data from these regions.
The resulting fits had $\chi^2_\nu\approx1$ and the
temperature and abundance distributions are shown in
Figures 2a \& b and Table 2. 
The temperature rises from $3.24_{ -0.06}^{+0.07}$ keV  within 2$^\prime$ to
$4.40_{-0.13}^{+0.13}$ keV beyond 5$^\prime$.
The central abundance is 0.53$^{+0.04}_{-0.04}$ solar, falling to
0.36$^{+0.03}_{-0.03}$ in the outer 3--12$^\prime$ region. 
In this combined data set, the central abundance enhancement within 
$3^\prime$ is significant at a confidence level of $\sim$99\%. 
Confidence contours for the abundances in the two innermost regions are shown
in Figure 3. 
We also used the $F$-test to assess the significance of the central
abundance enhancement. We simultaneously fitted the spectra from the
inner and outer regions, allowing their normalizations, temperatures
and abundances to vary independently.  We then refit the spectra with
the abundances from the two regions tied together.  The $\chi^2$ 
of these latter fits were larger by $\ge26$ for an increase of only one 
degree of freedom.  The $F$-test indicates that the central abundance 
enhancement is significant at a level of $>$99.99\%.

Since there is a moderate cooling flow at the center of Abell 496, we tested 
whether 
the abundance gradient described above is an artifact of our choice
of spectral model.
We added a cooling flow component to the {\tt mekal} thermal emission 
model in the central region.
The cooling flow spectral model {\tt cflow} in XSPEC is characterized by 
maximum and minimum temperatures, an abundance, a slope which parameterizes 
the temperature distribution of emission measures, and a 
normalization, which is simply the cooling accretion rate.
We adopted the emission measure temperature distribution that corresponds
to isobaric cooling flows (zero slope).
We tied the maximum temperature of the cooling flow to the temperature of the
thermal component, and we fixed the minimum temperature at 0.1 keV. 
We applied a single (but variable) global absorption to both spectral 
components and associated an additional, intrinsic absorption component with
the cooling flow, placing it at the redshift of the cluster.
The addition of the cooling flow component does not significantly affect our 
results for the central region:
the central abundance enhancement remains significant at a confidence
level of $>$90\%. 
The two component fits at the center are slightly worse 
than the isothermal model fits above, but they still have $\chi^2_\nu\approx1$ 
(see Table 2).
In order to apply the $F$-test to these cooling flow model fits,
we also simultaneously fit the spectra from inner and outer regions, first
allowing the abundances in the respective regions to vary independently,
then tying the abundances together.  The $F$-test implies that the 
abundance gradient is significant at the $>$99.99\% level.
Since the spectra we analyze are from cylindrical projections of 
emission measure through various lines of sight through the cluster,
the true spatial abundance enhancement at the center will be somewhat stronger
than we observe.

				\subsection{
Individual Elemental Abundances \& Abundance Ratios
                                }
 
We also determined the abundances of individual elements using
the {\tt vmekal} spectral model in XSPEC.
A similar analysis for the outer regions ($>3'$) of Abell 496 was done by 
Mushotzky et al.\ (1996), whose individual elemental abundance measures are  
consistent with our results at the 90\% confidence level.
In our spectral model fits, 
the He abundance was fixed at the solar value, while 
C and N were fixed at 0.3 solar (since {\sl ASCA} is rather insensitive
to C and N and the derived abundances of other elements 
are not affected by the particular choice for C and N abundances).
Our observed abundances are shown in  Table 3
for various projected spatial regions. 
  
Table 3 shows that the iron abundance is best determined and increases 
$\sim$50\% from $0.36\pm0.03$ solar in the outer parts ($>3^\prime$)
to a central value of $0.53\pm0.05$ solar.
The sulfur abundance also shows a significant gradient, rising from 
0.20$\pm0.16$ solar beyond 3$^\prime$ to 0.58$\pm0.20$ solar at the center.
The silicon abundance is $\sim$0.8$\pm0.2$ solar, showing no significant
gradient within its 90\% confidence limits. 
The best fitting neon abundance is also nearly solar, showing no significant
gradient, while the best fitting nickel abundance is $\sim$2.5 times solar at 
the center, with a marginally significant decline to solar in the outer parts.
Oxygen, magnesium, argon and calcium are poorly constrained.

Theoretical numerical models of supernova yields predict the following
elemental ratios relative to solar values: 
for SN Ia, the W7 model of Nomoto, Thielemann \& Yokoi (1984), as updated in
Nomoto et al. (1997a), gives
$$
{\rm O} \approx  {\rm Mg} \approx  0.035~ {\rm Fe},
$$
$$
{\rm Ne} \approx  0.006~{\rm Fe},
$$
$$
{\rm Si} \approx  {\rm S} \approx  {\rm Ar} \approx  {\rm Ca} \approx 0.5~
{\rm Fe},
$$
$$
{\rm Ni} \approx  4.8~{\rm Fe};
$$
while for  SN  II, Nomoto et al.\ (1997b) find
$$
{\rm O} \approx  {\rm Mg} \approx  {\rm Si} \approx 3.7~{\rm Fe},
$$
$$
{\rm Ne} \approx {\rm S} \approx  2.5~{\rm Fe},
$$
$$
{\rm Ar} \approx  {\rm Ca} \approx  {\rm Ni} \approx 1.7~{\rm Fe},
$$
after integrating their yields over a Salpeter mass function with 
upper and lower mass limits of 10 and 50 $M_\odot$, respectively.
 
Various observed abundance ratios within inner and outer projected spatial 
regions
are shown in Table 4, along with the 
theoretical expectations for SN Ia and SN II ejecta; the errors associated 
with the observed abundance ratios are the propagated $1\sigma$ errors.
Note that several abundance ratios lie significantly outside their theoretical 
ranges: Ne/Si ($0-3^\prime$), Ne/S ($3-12^\prime$) and Si/S 
($3-12^\prime$).
Three abundance ratios show marginally significant gradients: 
Si/S, Si/Ni, and S/Fe.

				\section{
Distinguishing the Relative Contributions from SN Ia \& SN II
                                }

We use the observed abundance ratios (normalized by their solar
values) in Table 4 to estimate the relative contribution of SN Ia and SN II to
the metal enrichment of the intracluster gas.
Such estimates are complicated by uncertainties
in both the observations and the theoretical yields.
The yield relations above show that the most discriminatory 
abundance ratios are the ones involving oxygen, magnesium and neon, since their
ratios to iron are 2-3 orders of magnitude smaller for SN Ia than for SN II; 
of these, magnesium is poorly determined, which leaves oxygen and neon.
Despite their large fractional uncertainties, the fact that the observed 
O/Fe and Ne/Fe ratios are of order unity (see Table 4) clearly indicates the
presence of SN II ejecta --- these ratios are predicted to be less than a few 
percent for pure SN Ia ejecta.
The 90\% errors in these abundance ratios are not so large that they 
can be consistent with SN Ia ejecta alone.
On the other hand, the observed O/Fe and Ne/Fe ratios are about
half the values predicted for pure SN II ejecta, which may
indicate dilution by SN Ia ejecta. The SN Ia iron mass fractions
indicated by these abundance ratios are indicated in Table 4.
However, the theoretical iron yields from SN II 
are uncertain by a factor of $\sim$2 (Woosley \& Weaver 1995; Arnett 1996; 
Gibson et al. \ 1997),
so this systematic uncertainty obscures the relative contribution from
SN Ia and SN II. 

The abundance ratios with the smallest fractional errors are Si/Fe, S/Fe 
and Ni/Fe. 
The Si/Fe and Ni/Fe ratios in Table 4 indicate a roughly comparable mix of 
SN II and SN Ia ejecta in the outer parts, while the SN Ia/II mix
indicated by the S/Fe ratio is inconsistent with those derived from
the Si/Fe and Ni/Fe ratios (we show below that elemental ratios involving
sulfur are problematic).
The associated SN Ia iron mass fractions are listed in Table 4 for these 
ratios as well.  
These three abundance ratios also involve iron, the production of which 
in SN II models is uncertain by a factor of $\sim$2, so we next consider 
SN Ia/II discriminators that do not involve iron. 

Two well-determined ratios independent of iron are Si/Ni and Si/S. 
The SN Ia/II ratio derived from Si/Ni is consistent with
the values derived above (see Table 4),
indicating that SN Ia contribute 74\% (66\%) of the
iron mass in the inner $0-2^\prime$ ($0-3^\prime$) region and
48\% in the outer $3-12^\prime$ region.
However, the best-fit value of the Si/S ratio is outside the theoretical 
boundaries in the outer parts of Abell 496, although its errors are large enough 
to be consistent with the expectation for SN II ejecta.
Inspection of Table 4 shows that most of the best fit values of the other 
ratios involving sulfur in the outer parts are also systematically 
outside the theoretical range.
We conclude, despite the large fractional errors, that
sulfur is likely to be overproduced in the SN II models we have adopted.  
To be consistent with the results from the other elemental ratios above, 
sulfur production in SNe II should be reduced by a factor of $\sim$2-4
relative to the models of Nomoto et al.\ (1997b).
This overproduction of sulfur in other SN II models was noted previously
by Mushotzky et al.\ (1996), Loewenstein \& Mushotzky (1996) and
Gibson et al. \ (1997).  Recently, Nagataki \& Sato (1998) 
found that this sulfur discrepancy was reduced when they used the theoretical
SN II yields of Nagataki et al.\ (1998), who explored the affects of
asymmetric SN II explosions.

Figure 4 summarizes our estimates of the iron mass fraction from SN Ia,
as derived from the variety of elemental abundance ratios described above; 
values for the inner $0-2^\prime$ region are indicated by filled circles, 
while empty circles correspond to $3-12^\prime$.
The filled and empty symbols are obviously segregated, despite the large
individual errors, indicating that the outer region is more dominated by 
SN II ejecta and the central region is dominated by SN Ia ejecta.
The consistency of the results for 
ratios $not$ involving sulfur is particularly noteworthy; 
in Figure 4 the average theoretical yield of sulfur 
from SN II has been reduced by a factor of 3.8 to bring the estimates from
sulfur ratios more in line with the other estimates.

For our consensus estimate of the iron mass fraction due to SN Ia, 
we average the SN Ia iron mass fractions for five of the seven
abundance ratios discussed 
above: O/Fe, Ne/Fe, Si/Fe, Ni/Fe and Si/Ni
(the two ratios involving sulfur are excluded).
Our best estimates for the SN Ia iron mass fraction are $70\pm5$\% 
within $0-2^\prime$, $64\pm5$\% within $0-3^\prime$, and $50\pm6$\% 
within $3-12^\prime$ (these values are denoted as $\mu$ in Figure 4).  
Although the exact proportions of SN Ia and SN II ejecta are sensitive 
to our adopted theoretical SN II yields, this constitutes 
evidence for an increase in the proportion of iron 
from SN Ia at the center relative to the outer parts.
For our adopted yields, the proportion of iron from SN Ia is 
$\sim$50\% larger at the center than the outer parts.
This increased proportion of SN Ia ejecta is such that
the central iron abundance enhancement can be attributed wholly to SN Ia.
Combining our results for the gradients in both the iron abundance 
and the relative 
proportion of SN Ia and II, we find that the iron abundance due to SN II
is $A_{\rm Fe_{\rm II}}=0.18\pm0.06$ solar in the outer parts ($3-12'$) and 
$0.16\pm0.06$ solar
near the center ($0-2'$), exhibiting no significant gradient.
The iron abundance due to SN Ia is $A_{\rm Fe_{\rm Ia}}=0.18\pm0.06$ solar 
in the outer parts, increasing a factor of $\sim$2 to $0.38\pm0.07$ 
solar near the center (see Table 5).

While our estimate of the relative proportion of SN Ia/II ejecta depends
upon our adopted SN yield models, our qualitative conclusion that there
is a large fraction of iron from SN Ia is robust.
Fukazawa et al.\ (1998) found in {\sl ASCA} spectra of 40 clusters that
Si/Fe ratios are lower in cool clusters than in hot clusters,
indicating that the proportion of SN Ia ejecta is larger in cool clusters
than in hot clusters.
Davis, Mulchaey \& Mushotzky (1999) recently showed that this trend continues
in galaxy groups dominated by early-type galaxies.
Given that both SN Ia and SN II produce significant amounts of Si and Fe (although
in different proportions), there must be roughly comparable amounts of ejecta from 
both SN Types in typical clusters to produce the Si/Fe -- $kT$ trend.  
Fukazawa et al.\ (1998) conclude that the SN Ia/II  mixture 
in clusters cannot be as ambiguous as suggested by Gibson et al.\ (1997).

                           \section{
Distinguishing between Possible Metal Enrichment Mechanisms
                                }

We showed above that iron, nickel and sulfur abundances are centrally 
enhanced in Abell 496 and that gradients in various elemental ratios indicated 
that these central abundance enhancements are largely due to SN Ia ejecta.
Our more model dependent result is that $\sim$50\% of the iron
$3-12'$ from the cluster center comes from SN Ia. 
The existence of spatial gradients in abundance ratios implies
that the dominant metal injection mechanism near the
cluster center must be different than in the outer parts.
We will first distinguish between several possible mechanisms for
producing central abundance enhancements in the cluster.
Then we will assess the relative roles of winds and ram pressure stripping
as metal enrichment mechanisms for the bulk of the cluster.

				\subsection{
Mechanisms for Creating Central Abundance Enhancements
				}

There are several mechanisms that may cause central abundance enhancements 
in intracluster gas:
1) ram pressure stripping of the metal-rich gas in cluster galaxies 
by intracluster gas is more effective at the center, where the intracluster
gas density is highest (Nepveu 1981);
2) secular mass loss from the stars in central dominant galaxies may
accumulate near the cluster center (White et al.\ 1994);
3) if galaxies blew winds which were not thoroughly mixed in the intracluster
gas, metal abundances would decline outward from the center,
since the luminosity density of metal-injecting galaxies falls
more rapidly with radius than the intracluster gas density
(Koyama et al.\ 1991; White et al.\ 1994);
4) even if cluster galaxies blew winds which were generally well-mixed
in the intracluster gas, a central dominant galaxy's wind may 
be at least partially suppressed, by virtue 
of its location at the bottom of the cluster's gravitational potential
and in the midst of the highest intracluster gas density
(White et al.\ 1994).
Although each of these mechanisms can produce abundance gradients, no 
individual mechanism can produce spatial gradients in abundance $ratios$. 
The fact that we observe gradients in abundance ratios 
implies that the dominant metal enrichment mechanism changes spatially.
 
There are several reasons why ram pressure stripping is 
unlikely to be the source of SN Ia iron at the center of the cluster.
First, the gaseous abundances measured in most early-type galaxies by
{\sl ASCA} (Loewenstein et al.\ 1994; Matsumoto et al.\ 1997)
and {\sl ROSAT} (Davis \& White 1996) are 0.2-0.4 solar, 
significantly less than the 0.5-0.6 solar abundance observed at
the cluster center.
Only the most luminous ellipticals, which also tend to be at the
centers of galaxy clusters or groups, are observed to have gaseous 
abundances of 0.5-1 solar.
Second, if ram pressure stripping is the primary source of SN Ia material at the 
center of the cluster, the cD is the one galaxy in the cluster which should 
not be stripped.  Therefore, the cD should exhibit its accumulated history of
SN Ia ejecta, in addition to any ejecta stripped from other galaxies
as they passed near the cluster center.
The cD should then have more SN Ia iron than expected for its luminosity.
However, we show in Appendix A that the SN Ia iron mass to luminosity ratio 
($M_{ {\rm Fe}_{\rm Ia} }/L_{{B_{ {\rm E}/{\rm S0} } }}$) in the vicinity 
of the cD is no 
greater than in the rest of the cluster (see Table 5).  
This indicates that the bulk of the SN Ia ejecta from the cD has been
retained in its vicinity, but has not been supplemented by ejecta
from other galaxies. Third, gaseous abundances in ellipticals tend to decline 
outward from their centers, as observed in NGC 4636 (Matsushita et al.\ 1997)
and other early-type galaxies (Matsushita 1997),
since the mass-losing stars in ellipticals exhibit such metallicity gradients;
thus, most stripped gas will have even lower abundances than indicated
by the global X-ray measures of ellipticals
(since global measures are weighted by centrally concentrated
emission measures in ellipticals), which are already too low. 
Finally, since the efficiency of ram pressure stripping depends on the
intracluster gas density, which declines strongly with radius,
the abundances of SN Ia ejecta should also decline strongly with radius,
which is not observed; we see only a factor of two decline in the iron 
abundance from SN Ia (see Table 5). 
We conclude that ram pressure stripping is not the cause of the
central abundance enhancements in Abell 496.
If ram pressure stripping is effective in the cluster, it would act
to $dilute$ the central abundance enhancement.

To assess whether secularly accumulated stellar mass loss in the cD
could cause the observed central abundance enhancement in Abell 496,
we compare the cD to a giant elliptical, which is not in a rich cluster:
NGC 4636 is one of the most X-ray luminous ellipticals 
and may be at the center of its own small group (Matsushita et al.\ 1998).
Thus, NGC 4636 is not in an environment where it is likely to have
been stripped.  If giant (unstripped) ellipticals secularly accumulate 
their stellar mass loss (after a putative wind phase),
NGC 4636 should have the same SN Ia iron mass to 
light ratio as the cD in Abell 496.
Instead, we find that the SN Ia Fe mass to light ratio in the central region 
of Abell 496 is at least $\sim$ 20 times higher than in NGC 4636 (see
Appendix A).
We conclude that the central abundance enhancements in Abell 496 are not 
likely to be due to secularly accumulated stellar mass loss in the cD.

If early-type galaxies blew winds which were only locally mixed in
the intracluster gas, we would expect to observe an abundance gradient $A(r)$
which is proportional to the ratio of the luminosity density of wind-blowing
galaxies ($\ell_{{\rm E}/{\rm S0}}$) to the intracluster gas density, i.e. 
$A(r)\propto \ell_{{\rm E}/{\rm S0}}(r)/\rho_{\rm gas}(r)$.
Equivalently, 
$A(r)\propto L_{{\rm E}/{\rm S0}}(r)/M_{\rm gas}(r) $, 
where $ L_{{\rm E}/{\rm S0}}(r)$
and $M_{\rm gas}(r)$ are the luminosity of early-type galaxies
and the intracluster gas mass enclosed in spherical shells about the center.
We derived the luminosity distribution of early-type galaxies in
Abell 496 from the galaxy morphology data of Dressler (1980) and
the gas density distribution from $Einstein$ IPC data (see Appendix A).
In Table 5 we show that the ratio $L_{{\rm E}/{\rm S0}}/M_{\rm gas}$
declines a factor of $\sim4$ from $0-2'$ to $3-12'$, while the observed
iron abundance declines only $30\%$ (and the SN Ia iron abundance
drops by half).  
Thus, the observed abundance gradient is too shallow to be caused by poorly
mixed winds from early-type galaxies in the cluster.

We propose instead that the central gradients in abundances and abundance ratios 
in Abell 496 result from a partially suppressed SN Ia-driven wind from the cD.
This would be a secondary wind phase, following a more vigorous, unsuppressed
protogalactic wind driven by SN II.
Such a secondary wind phase in noncentral early-type galaxies can  
generate the SN Ia enrichment seen in the bulk of the intracluster gas, 
as we suggest in the next subsection.
As we showed above, the SN II iron mass to light ratio in the vicinity of
the cD and the lack of a central enhancement in SN II ejecta
indicate that it lost the bulk of its SN II ejecta.
However, a weaker SN Ia-driven wind would be more readily suppressed at 
center of the cluster, due to the depth of the gravitational potential
and the high ambient intracluster gas density.
SN Ia-driven winds would be less vigorous than the initial
SN II-driven winds, since SN Ia inject $\sim$10 times less energy per unit 
iron mass than SN II;
the observations indicate that comparable amounts
of iron came from SN Ia and SN II, so SN II have injected 
$\sim$10 times more energy into the intracluster gas than SN Ia.

If suppressed winds from central dominant galaxies 
are the cause of central abundance enhancements in 
other clusters, the prevalence of such enhancements in cooler clusters 
may be related to their cooling flow properties.
Cool clusters tend to have cooling flows with smaller accretion 
rates than hot clusters, so the history of prior metal ejection
from the central cD may be more likely to survive in cool clusters.
For a given cD optical luminosity, the metal ejection is more likely to 
extend beyond the cooling flow region (since the central intracluster gas
density is smaller in cool clusters than hot clusters), and the inward 
advection of the cooling flows would be less destructive to preexisting 
abundance gradients than in hotter clusters with higher accretion rates.

				\subsection{
Global Enrichment Mechanisms
				}

As mentioned in the introduction, the two metal enrichment mechanisms 
usually considered for the bulk of intracluster gas are protogalactic 
winds from early-type galaxies and ram pressure stripping of gas from 
galaxies in the cluster.
The yields of the SN II models that we adopt
(Nomoto et al.\ 1997b) lead 
us to conclude that nearly 50\% of the iron in Abell 496 comes from SN Ia.
Similar conclusions were reached for other clusters by
Ishimaru \& Arimoto (1997) and Nagataki \& Sato (1998), who used different
theoretical models for SN II.
Slightly more than half of the cluster iron comes from SN II, which
can be readily attributed to protogalactic wind enrichment.
However, the quantity and spatial extent of the iron from SN Ia 
is problematic: is ram pressure stripping so effective that it 
contaminates the outer parts of clusters nearly as effectively as 
the central regions?
Or was there a (secondary) galactic wind phase driven by SN Ia?

If ram pressure stripping is effective in the outer
parts of the cluster, it should be even more effective at the
center, where the intracluster gas density is highest.
However, we showed in the previous subsection that ram pressure stripping 
cannot account for the central concentration of metals.
Renzini et al.\ (1993) have also argued strongly against ram pressure stripping
being very significant in clusters, citing the lack of a strong metallicity 
trend with cluster temperature: hot clusters have higher velocity dispersions
and tend to have much higher gas densities than cool clusters, so ram pressure
stripping should be much more effective in hot clusters than cool clusters.
The lack of a strong metallicity trend with cluster temperature implies that
ram pressure is not the major source of intracluster metals.

We propose instead that the bulk of intracluster gas is contaminated 
by two phases of winds from early-type galaxies:
an initial SN II-driven protogalactic wind phase, followed by a secondary,
less vigorous SN Ia-driven wind phase.
As mentioned in the previous subsection, secondary SN Ia-driven winds would 
be $\sim$ 10 times less energetic than the initial SN II-driven
protogalactic winds. 
Fukazawa et al.\ (1998) invoked SN II-driven protogalactic
winds to account for their discovery that the proportion of SN Ia ejecta
is higher in cool clusters than in hot clusters.
They suggested that protogalactic winds
were energetic enough that SN II-enriched material was able to escape cool 
clusters, which have shallower gravitational potentials than hot clusters.
Less vigorous secondary SN Ia-driven winds would allow SN Ia-enriched
material to escape most galaxies, but not clusters.

The two phase wind scenario we are advocating has not been explicitly 
modeled to date.
Some previous evolutionary models for the gas in ellipticals have 
presumed the existence of an initial SN II-driven wind, without modeling it, 
and concentrated on a later SN Ia-drive wind phase
(Loewenstein \& Mathews 1991; Ciotti et al.\ 1991).
Some of these models assume that the bulk of intracluster iron comes from SN Ia.
Others investigators have assumed that the bulk of intracluster iron comes 
from SN II and model only SN II-driven protogalactic winds in detail 
(Larson \& Dinerstein 1975; David, Forman, \& Jones 1991).
However, in all recent investigations, present epoch SN Ia rates were adopted 
which lead to huge overpredictions of the current gaseous abundances in 
ellipticals: iron abundances were theoretically predicted to be 3-5 times solar, 
while {\sl ASCA} and {\sl ROSAT} observations find abundances to be 0.1-1 solar.

In parlance similar to that of Renzini et al.\ (1993), we are proposing a
wind-outflow-inflow (WOI) model in which the ``wind" is driven by SN II,
the less vigorous ``outflow" is driven by SN Ia, and the subsequent
inflow experiences much less contamination by SN Ia than in previous modeling.
Generating SN Ia-enriched outflows that can contaminate intracluster 
gas, but leave current abundances subsolar in elliptical atmospheres, 
requires rather different SN Ia rate evolution than in previous models.
Prior models of SN Ia-driven winds tended to inject roughly the right
amount of iron in clusters (to within a factor of $\sim$2 or so).
However, the SN Ia rate must decline much faster than in previous models
if the current SN Ia rate is to be as low as $\lesssim0.03$ SNU 
(Loewenstein \& Mushotzky 1997), in order to match the low iron abundances in
elliptical atmospheres.
This rate is 3-10 times smaller than previously adopted and is $\sim4$
times smaller than the most recent optical estimate of the current 
SN Ia rate in ellipticals (0.13 SNU; Capellaro et al.\ 1997).
To generate the amount of SN Ia observed in intracluster gas,
the SN Ia rate at earlier times must be much larger than previously modeled,
to compensate for the much lower current rate.
These heuristic constraints on the evolution of the SN Ia rate are 
not yet theoretically motivated.

                               \section{
Summary  
                                }

We have carried out a detailed analysis of the distribution of elemental 
abundances in the intracluster gas of Abell 496. 
Our main results which are independent of our choice of supernovae yield models 
are:
\begin{enumerate}
\item The hot gas of Abell 496 has significant abundance gradients: the iron 
abundance 
 is 0.36$^{+0.03}_{-0.03}$ solar  3--12$^\prime$ from the center, rising
   $\sim$50\% to 0.53$^{+0.04}_{-0.04}$ solar within $2^\prime$;
   nickel and sulfur also have significant central concentrations. 
\item  There are spatial gradients in elemental abundance $ratios$ in this 
   cluster; a variety of abundance ratios individually and collectively indicate 
   that SN Ia
   ejecta is more dominant in the center than in the outer parts.
\item  We find no significant gradient in SN II ejecta.
\item  Ram pressure stripping is unlikely to generate the observed 
   central abundance
   enhancements for several reasons, including the fact that gaseous abundances observed 
   in elliptical atmospheres 
   tend to be substantially less than the abundances observed in the 
   intracluster gas near the center. 
\item  Two stage galactic winds, consisting of SN II-driven protogalactic 
   winds followed by 
   less energetic SN Ia-driven outflows, are proposed to generate comparable
   levels (by mass) of iron contamination from SN Ia and SN II in the 
   intracluster gas of Abell 496. 
\item    Since the secondary 
   SN Ia-driven wind is $\sim 10$ times less energetic than the SN II-driven wind, 
   it is more likely to be 
   smothered due to the cD being at the bottom of the 
   cluster's gravitational potential and in the midst of the highest 
   intracluster gas density; 
   Such a smothered wind  may generate the observed central
   SN Ia iron abundance enhancement.
   \end{enumerate}

Results which are more dependent upon our particular choice of supernovae yield
models include:
\begin{enumerate}
\item  SN Ia account for $\sim$50\% of the iron mass $3-12^\prime$ from 
   the center 
   of the cluster and $\sim$70\% of the iron mass within $2^\prime$.
\item  The central iron abundance enhancement can be attributed wholly to the 
   iron associated with the central enhancement of SN Ia ejecta.
\end{enumerate}

Our scheduled {\sl Chandra} observation of Abell 496 should allow us to 
trace more accurately the gradients in abundances and 
abundance ratios, given {\sl Chandra}'s higher spatial resolution and 
relative lack of scattering compared to {\sl ASCA}.
Since our suggested mechanism for generating gradients in abundances
and abundance ratios should not be specific to just one cluster, we
will be applying similar analyses to {\sl ASCA} data for other clusters,
as well.

			\acknowledgments
This work was partially supported by the NSF and the State of Alabama through
EPSCoR grant EHR-9108761.  REW also acknowledges partial support from NASA grant 
NAG 5-2574 and a National Research Council Senior Research Associateship
at NASA GSFC. 
This research made use of the HEASARC {\sl ASCA} database and NED.

\appendix
\section{
Appendix}

We calculate the iron mass to luminosity ratio in the vicinity of the cD
and for the rest of the cluster.
For the cD, we will separately consider the iron from SN Ia and SN II and 
we will treat the iron from SN Ia in two ways:
we will first attribute to the cD all the SN Ia iron in the central region
where the abundance is enhanced; however, this may be an overestimate,
since the cD's central location makes it difficult to separate general 
intracluster gas from its own interstellar medium, even if the latter 
is substantial. 
Consequently, we will also consider the possibility that only the ``excess" 
iron at the center (that which gives rise to the central abundance enhancement 
and to the central change in elemental abundance ratios) was generated by the
cD. 
This ``excess" is one third of all the iron within $3^\prime$,
or half of the iron attributable to SN Ia.

We derived the intracluster gas density distribution in Abell 496
by using {\sl Einstein} IPC data to determine the shape (core radius 
and asymptotic slope of a $\beta$-model) of its 
X-ray surface brightness distribution and using the {\sl ASCA}
observations described in \S2 to provide the flux normalization.
Given the iron abundances listed in Table 3, we calculated the iron mass
within the spherical volume contained between 
$0-3^\prime$  to be $M_{\rm Fe}\approx 5.1 \times 10^{9} M_{\odot}$, 
as listed in Table 5 (also listed are calculations for $0-2'$).
Using our best estimate of the SN Ia iron mass fraction described in
\S4, we calculated the iron mass from SN Ia to be 
$M_{\rm Fe_{\rm Ia}}\approx 3.3 \times 10^{9} M_{\odot}$ within $0-3'$.

We derived the galaxies' optical luminosity distribution in Abell 496 from the 
galaxy morphological and ($V$-band) photometric data of Dressler (1980).
We assumed that early-type galaxies are the source of the intracluster iron 
(cf.\ Arnaud et al.\ 1992) and derived their cumulative spherical luminosity 
distribution from a deprojection of their cumulative surface luminosity 
distribution. 
We converted the Dressler (1980) $V$ magnitudes to $B$ by assuming 
$B-V=1$ for early-type galaxies.
For the central cD, we used the photometry of Valentijn (1983),
which assigns a total $B$ magnitude of $B_T=13.42$, giving
a blue luminosity of $L_B=2.6\times10^{11}h_{50}^{-2}$ $L_\odot$.
We distributed the luminosity of the cD over several radial bins,
using an $r^{1/4}$ law with the same effective radius, 
to avoid an artificial luminosity spike at the center. 
Table 5 lists the iron mass to optical light ratio for SN Ia and II ejecta 
within various projected regions, 0--2$^{\prime}$, 0--3$^{\prime}$ and 
3--12$^{\prime}$. 
The errors shown in Table 5 include the iron mass and luminosity errors; 
the fitting errors from the luminosity deprojection 
procedure are relatively small, so are not included.

It can be seen from Table 5 that the iron mass to light ratio for SN II 
ejecta is $\sim$2-10 times smaller at the center than in the outer regions.
If ellipticals had protogalactic winds driven by SN II, this shows that
such a wind in the cD was not suppressed by it being at the cluster center 
(i.e. by being at the bottom of the gravitational potential well of the 
cluster and in the midst of the highest intracluster gas density).
Given that we see no significant gradient in SN II ejecta (see \S4), this 
suggests that the SN II ejecta in vicinity of the cD is simply the result
of a fairly uniformly mixed contamination in the cluster (in the regions 
observed).

The nominal SN Ia iron mass to light ratio in the central region is somewhat 
less than that of the outer parts, but the associated errors are large enough
that this difference is not significant. 
Thus, the bulk of the SN Ia ejecta produced by the cD has been retained in
its vicinity.

We also use X-ray observations of an individual elliptical galaxy to see how 
much gaseous iron it has accumulated per unit optical luminosity and
compare with the cD in Abell 496.
NGC 4636 is a particularly X-ray luminous elliptical for its optical 
luminosity, $10^\circ$ from the center of the Virgo cluster in the 
Virgo Southern Extension (Nolthenius 1993).
Its metric X-ray to optical luminosity ratio is $\sim$5 times larger
than the median for ellipticals (White \& Davis 1997, 1999), indicating that 
it has been particularly successful in retaining its hot gas. 
Matsushita et al.\ (1998) suggest that NGC 4636 may even be at the center 
of a small group of galaxies, with its emission enhanced by group gas.
If this is true, it makes our conclusion even stronger.
The optical and X-ray luminosities of NGC 4636 are 
$L_B \approx 3 \times 10^{10} L_{\odot}$ and 
$L_X \approx 4 \times 10^{41}$ erg s$^{-1}$, adopting a distance of 17 Mpc.
Very deep {\sl ASCA} exposures of this galaxy have been analyzed by 
Matsushita et al.\ (1997; 1998), who found an abundance gradient
characterized by a central value of $\sim$0.65 solar (converted to the 
photospheric abundance scale), declining to $\sim$0.2 solar $10'$ away.

We used the gas distribution of Matsushita et al.\ (1998) and the abundance
distribution of Matsushita et al.\ (1997) to calculate the iron mass
within $7r_{\rm eff}$ in NGC 4636.
This encompasses the bulk of its iron content and  virtually all of its
optical light.
We compare this to the iron mass and optical luminosity within $3'$ of the 
center of Abell 496, which encompasses most of the 
region with enhanced abundances.
We find that the SN Ia iron mass to luminosity ratio for the cD is 
$\sim$40 times greater than in NGC 4636.
If we attribute to the cD only the ``excess" amount of SN Ia iron at the center
compared to the rest of the cluster, the SN Ia iron mass to light ratio
in the cD is still $\sim$20 times higher than within NGC 4636.
The discrepancy is actually even greater than indicated, since we have 
overestimated the SN Ia iron within NGC 4636 by assuming that all of 
its iron was produced by SN Ia and we have underestimated the SN Ia
iron in the vicinity of the cD by restricting ourselves to within
$3'$, while the SIS data show that abundances are enhanced out to $\sim5'$.

\clearpage
 
				
                                \clearpage

                                \begin{figure}
                                \title{
Figure Captions
                                }
				\figcaption{
a) GIS 2 \& 3 spectra 
for inner ($0 - 2'$) ($top$) and outer ($3 - 12'$) ($bottom$) regions;
solid lines represent the best-fitting isothermal model; 
individual lines and line complexes are identified;
b) SIS 0 \& 1 spectra 
for inner ($0 - 2'$) ($top$) and outer ($3 - 12'$) ($bottom$) regions;
solid lines represent the best-fitting isothermal model.
                                }

                                \figcaption{
Results of simultaneous isothermal fits to SIS 0 \& 1 and GIS 2 \& 3 data.  
Errors are 90\% confidence limits; $1'\approx57 h_{50}^{-1}$.
a) Temperature (in keV) distribution;
b) Abundance (relative to solar) distribution.
                                }
 
				\figcaption{
Confidence contours for two interesting parameters;
abundances and temperatures from simultaneous fits of the two innermost 
($0 - 3'$ \& $3 - 6'$) regions. 
The three contours correspond to 68\%, 90\% and 99\% confidence limits.
                                }

				\figcaption{
Comparison of derived SN Ia Fe mass fractions obtained from various observed 
elemental 
abundance ratios for inner ($0-2^\prime$) and outer ($3-12^\prime$) regions. 
The ratios involving sulfur have been corrected by reducing the 
theoretical SN II sulfur production by a factor of 3.8. 
The average SN Ia mass fraction derived in this work is denoted $\mu$.
				}

                                \end{figure}

\clearpage

\begin{deluxetable}{lcc}
\small
\tablewidth{0pt}
\tablecaption{Effective Exposure Times}
\tablehead{
\colhead{Spectrometer } &
\colhead{Exposure Time  } \nl
\colhead{} &
\colhead{(kiloseconds)}  &
}
\startdata
SIS 0 & 29.6\nl
SIS 1 & 23.3\nl
GIS 2 & 39.8\nl
GIS 3 & 39.6\nl
\enddata
\end{deluxetable}
 
\begin{deluxetable}{cccc}
\small
\tablewidth{0pt}
\tablecaption{Joint Spectral Fits\tablenotemark{a}}
\tablehead{
\colhead{Region \tablenotemark{b}}  &
\colhead{$kT$}  &
\colhead{Abundance} &
\colhead{$\chi^2_\nu$}  \nl
\colhead{(arcmin)} &
\colhead{(keV)}  &
\colhead{(solar)} &
\colhead{}  
}
\startdata
$0 - 3$ & 3.40$^{+0.06}_{-0.06}$ & 0.49$^{+0.03}_{-0.03}$ & 1.07\nl
 $0 - 3^{c}$ & 3.70$^{+0.01}_{-0.08}$ & 0.51$^{+0.04}_{-0.03}$ & 1.06\nl
 $3 - 6$ & 4.17$^{+0.10}_{-0.09}$ & 0.37$^{+0.04}_{-0.04}$ & 1.04\nl
 $6 - 9$ & 4.46$^{+0.16}_{-0.16}$ & 0.32$^{+0.05}_{-0.05}$ & 1.07\nl
 $9 - 12$ & 4.44$^{+0.26}_{-0.24}$ & 0.42$^{+0.09}_{-0.09}$ & 1.02\nl
 $6 - 12$ & 4.40$^{+0.13}_{-0.13}$ & 0.34$^{+0.05}_{-0.04}$ & 1.11\nl
 $3 - 12$ & 4.28$^{+0.08}_{-0.08}$ & 0.36$^{+0.03}_{-0.03}$ & 1.07\nl
 $0 - 2$ & 3.24$^{+0.07}_{-0.06}$ & 0.53$^{+0.04}_{-0.04}$ & 1.03\nl
 $0 - 2^{c}$ & 3.37$^{+0.08}_{-0.09}$ & 0.54$^{+0.05}_{-0.05}$ & 1.03
\enddata
\tablenotetext{a}{Errors are 90\% confidence limits}
\tablenotetext{b}{Distance from the X-ray center }
\tablenotetext{c}{same as above but with an extra  cooling flow component}
\end{deluxetable}
 
\begin{deluxetable}{lcccc}
\small
\tablewidth{0pt}
\tablecaption{Individual Elemental Abundances\tablenotemark{a}}
\tablehead{
\colhead{} &
\colhead{} &
\colhead{Region} &
\colhead{} &
\colhead{}\nl
\colhead{Element} &
\colhead{$0 - 2'$ }  &
\colhead{$0 - 3'$ } &
\colhead{$3 - 12'$ } &
\colhead{gradient?}
}
\startdata
O & 0.48$^{+0.39}_{-0.37}$ & 0.62$^{+0.33}_{-0.32}$ & 0.57$^{+0.47}_{-0.29}$ \nl
Ne & 0.87$^{+0.45}_{-0.40}$ & 0.89$^{+0.36}_{-0.33}$ & 0.73$^{+0.48}_{-0.28}$  
\nl
Mg & 0.20$^{+0.34}_{-0.20}$ & 0.33$^{+0.27}_{-0.29}$  & 0.01$^{+0.35}_{-0.01}$ 
\nl
Si & 0.83$^{+0.18}_{-0.17}$ & 0.83$^{+0.12}_{-0.14}$ & 0.69$^{+0.19}_{-0.13}$ 
\nl
S & 0.58$^{+0.19}_{-0.20}$ & 0.49$^{+0.16}_{-0.16}$ & 0.20$^{+0.15}_{-0.17}$ & 
$\surd$\nl
Ar & $\le$0.3 & $\le$0.3 & 0.01$^{+0.45}_{-0.01}$ \nl
Ca & 0.33$^{+0.48}_{-0.33}$ & 0.22$^{+0.61}_{-0.22}$ & 0.31$^{+0.49}_{-0.31}$ 
\nl
Fe & 0.53$^{+0.05}_{-0.05}$ & 0.50$^{+0.04}_{-0.04}$ & 0.36$^{+0.03}_{-0.03}$ & 
$\surd$ \nl
Ni & 2.57$^{+0.67}_{-0.80}$ & 2.01$^{+0.58}_{-0.64}$ & 1.04$^{+0.79}_{-0.60}$ & 
$\surd$ 
\enddata
\tablenotetext{a}{Errors are 90\% confidence limits}
\end{deluxetable}

\begin{deluxetable}{lccccccccc}
\small
\tablewidth{0pt}
\tablecaption{Elemental Abundance Ratios\tablenotemark{a}}
\tablehead{
\colhead{Element} &
\colhead{grad} &
\colhead{} &
\colhead{Region} &
\colhead{} &
\multicolumn{2}{c} {Theory\tablenotemark{b}} &
\multicolumn{3}{c}{Fe Mass Fraction from SN Ia} \nl
\colhead{Ratio} &
\colhead{} &
\colhead{$0 - 2'$ }  &
\colhead{$0 - 3'$ } &
\colhead{$3 - 12'$ } &
\colhead{SN Ia} &
\colhead{SN II} &
\colhead{$0 - 2'$} &
\colhead{$0 - 3'$} &
\colhead{$3 - 12'$}
}
\startdata
O/Fe & & 0.91$^{+0.44}_{-0.44}$ & 1.23$^{+0.39}_{-0.39}$ & 1.58$^{+0.62}_{-
0.60}$ & 0.037 & 3.82 
& 0.77$^{+0.12}_{-0.13}$ & 0.68$^{+0.11}_{-0.10}$ & 0.59$^{+0.16}_{-0.16}$  \nl
Ne/Fe & & 1.64$^{+0.53}_{-0.45}$ & 1.77$^{+0.66}_{-0.38}$ & 2.00$^{+0.66}_{-
0.38}$ &  0.006 & 2.69 
& 0.39$^{+0.17}_{-0.20}$ & 0.34$^{+0.14}_{-0.24}$ & 0.26$^{+0.14}_{-0.20}$  \nl
Si/Fe & & 1.58$^{+0.22}_{-0.22}$ & 1.68$^{+0.21}_{-0.17}$ & 1.92$^{+0.29}_{-
0.27}$ &  0.538 & 3.53 
& 0.65$^{+0.08}_{-0.07}$ & 0.62$^{+0.06}_{-0.07}$ & 0.54$^{+0.09}_{-0.10}$  \nl
Ni/Fe & & 4.86$^{+1.02}_{-0.96}$ & 4.01$^{+0.84}_{-0.78}$ & 2.88$^{+1.12}_{-
1.05}$ &  4.758 & 1.65 
& 1.00$^{+0.00}_{-0.27}$ & 0.76$^{+0.24}_{-0.25}$ & 0.40$^{+0.36}_{-0.34}$  \nl
O/Si & & 0.58$^{+0.72}_{-0.29}$ & 0.73$^{+0.58}_{-0.24}$ & 0.82$^{+1.47}_{-
0.33}$ & 0.068 & 1.1 
& 0.87$^{+0.09}_{-0.86}$ & 0.78$^{+0.12}_{-0.78}$ & 0.69$^{+0.21}_{-0.69}$  \nl
Si/Ni & $\surd$& 0.33$^{+0.08}_{-0.08}$ & 0.42$^{+0.1}_{-0.09}$ & 
0.67$^{+0.27}_{-0.26}$ &  0.113 & 2.14 
& 0.74$^{+0.09}_{-0.07}$ & 0.66$^{+0.08}_{-0.08}$ & 0.48$^{+0.19}_{-0.14}$  \nl
Ne/Si\tablenotemark{c} & & 1.04$^{+0.36}_{-0.31}$ & 1.05$^{+0.41}_{-0.24}$ & 
1.05$^{+0.35}_{-0.31}$ & 0.012 & 0.76
& & \nl
S/Fe & $\surd$ & 1.10$^{+0.19}_{-0.25}$ & 0.98$^{+0.22}_{-0.28}$ & 
0.54$^{+0.30}_{-0.29}$ &  0.585 & 2.29 
& & \nl
O/S & & 0.82$^{+0.42}_{-0.43}$ & 1.26$^{+0.49}_{-0.53}$ & 2.92$^{+1.98}_{-1.85}$ 
&  0.063 & 1.67 
& & \nl 
Ne/S\tablenotemark{c}& & 1.48$^{+0.53}_{-0.52}$ & 1.81$^{+0.79}_{-0.64}$ & 
3.72$^{+2.35}_{-2.16}$ & 0.001 & 1.18 
& & \nl
Si/S\tablenotemark{c} & $\surd$& 1.43$^{+0.30}_{-0.37}$ & 1.72$^{+0.43}_{-0.52}$ 
& 3.55$^{+2.03}_{-1.88}$ &  0.919 & 1.54 
& & \nl
S/Ni & & 0.23$^{+0.06}_{-0.07}$ & 0.24$^{+0.07}_{-0.08}$ & 0.19$^{+0.13}_{-
0.12}$ &  0.123 & 1.39 
& & 
\enddata
\tablenotetext{a}{Errors are  propagated 1$\sigma$ errors}
\tablenotetext{b}{SN Ia: Nomoto et al (1997a); SN II: Nomoto et al (1997b)}
\tablenotetext{c}{Significantly out of theoretical boundaries}
\end{deluxetable}

\begin{deluxetable}{lccccc}
\small
\tablewidth{0pt}
\tablecaption{ Gas \& Iron Masses from SN Ia \& II }
\tablehead{
\colhead{} &
\colhead{} &
\colhead{Region} &
\colhead{} & \nl
\colhead{} &
\colhead{$0-2^{\prime}$} &
\colhead{$0-3^{\prime}$}  &
\colhead{$3-12^{\prime}$}  &
}
\startdata
$M_{\rm Fe_{\rm Ia}}$ ($M_\odot$)& $(1.7\pm 0.3) \times 10^{9}$ & $(3.3\pm 0.6) 
\times 
10^{9}$ & $(1.6\pm 0.3) \times 10^{10}$ \nl
$M_{\rm Fe_{\rm II}}$ ($M_\odot$)& $(0.7\pm 0.2) \times 10^{9}$  & $(1.8\pm 0.4) 
\times 
10^{9}$ & $(1.6\pm 0.3) \times 10^{10}$  \nl
${M_{\rm Fe_{\rm Ia}}}/{M_{\rm Fe}} $ & $0.70\pm 0.05$ & $0.64\pm 0.05$ & 
$0.50\pm 0.06$  \nl
$A_{\rm Fe_{\rm Ia}}$ (solar) & $0.38\pm0.07$ & $0.33\pm0.07$ & $0.18\pm0.06$ 
\nl
$A_{\rm Fe_{\rm II}}$ (solar) & $0.16\pm0.06$ & $0.18\pm0.06$ & $0.18\pm0.06$ 
\nl
${M_{\rm Fe_{\rm Ia}}}/{L_{B_{{\rm E}/{\rm S0}}}} $ & $0.015\pm 0.009$ & 
   $0.023\pm 0.015$ & $0.033\pm 0.020$ \nl
${M_{\rm Fe_{\rm II}}}/{L_{B_{{\rm E}/{\rm S0}}}} $ &  $0.006\pm 0.004$ & 
   $0.013\pm 0.008$ & $0.033\pm 0.020$ \nl
$M_{\rm gas}$ ($M_\odot$)& $2.2\times10^{12}$ & 
$5.1\times10^{12}$&$4.5\times10^{13}$\nl
$L_{B_{{\rm E}/{\rm S0}}}$ & $1.1\times10^{11}$ & 
$1.4\times11^{11}$ & $4.9\times10^{11}$  \nl
$L_{B_{{\rm E}/{\rm S0}}}/M_{\rm gas}$ & 0.050 & 0.028 & 0.011 \nl
\enddata
\end{deluxetable}


\begin{references}

                                \reference{}
Anders, E. \& Grevesse N. 1989, Geochimica et Cosmochimica Acta, 53, 197
 
				\reference{}
Arnaud, K. A. 1996, in {\it Astronomical Data Analysis Software and Systems V}, 
    ASP Conf. Series volume 101, eds. Jacoby, G. \& Barnes, J., p.17 

				\reference{}
Arnaud, K. A., Mushotzky, R. F., Ezawa, H., Fukazawa, Y., Ohashi, T., 
    Bautz, M. W., Crewe, G. B., Gendreau, K. C., Yamashita, K., Kamata, Y., 
    Akimoto, F. 1994, \apjl, 436, L67

                                \reference{}
Arnaud, M., Rothenflug, R., Boulade, O., Vigroux, L. \& Vangioni-Flam, E. 1992,
        \aap, 254, 49

                                \reference{}
Arnett, D. 1996, {\it Supernovae and Nucleosynthesis} 
   (Princeton: Princeton Univ. Press)

                                \reference{}
Bird, C. M. 1993, PhD thesis, Minnesota university, Minneapolis

				\reference{}
Canizares, C. R., Markert, T. H., \& Donahue, M. E. 1988, in {\it Cooling flows 
    in clusters and galaxies}, ed. A. C. Fabian (Dordrecht: Kluwer), p. 63

				\reference{}
Canizares, C. R.; Clark, G. W.; Jernigan, J. G.; Markert, T. H. 1982, \apj, 262, 
33

				\reference{}
Capellaro, E., Turatto, M., Tsvetkov, D. Yu., Barttunov, O. S., 
   Pollas, C., Evans, R., Hamuy, M. 1997, \aap, 322, 431

				\reference{}
Ciotti, L., Pellegrini, S., Renzini, A.,  D'Ercole, A. 1991, \apj, 376, 380
 
                                \reference{}
Cowie, L.  L., Hu, E .M., Jenkins, E. B. \& York, D. G. 1983, \apj, 272, 29

                                \reference{}
Dickey, J. \& Lockman, F. J. 1990, \araa, 28, 215

				\reference{}
David, L. P., Forman, W., \& Jones, C. 1991, \apj, 380, 29

				\reference{}
David, L. P., Forman, W., \& Jones, C. 1991, \apj, 359, 39

				\reference{}
Davis, D. S., Mulchaey, J. S. \& Mushotzky, R. F.  1999, \apj, 511, 34

				\reference{}
Davis, D. S. \& White, R. E. III 1996, \apjl, 470, L35

                                \reference{}
Dressler, A. 1980, \apjs, 42, 565

                              \reference{}
Dupke, R. A., 1998, PhD Thesis, University of Alabama, Tuscaloosa

                                 \reference{}
Dupke, R. A., \& White, R. E. III, 2000, \apj, 528, 139

				\reference{}
Ezawa, H., Fukazawa, Y., Makishima, K., Ohashi, T., Takahara, F., Xu, H.,
    Yamasaki, N. Y. 1997, \apjl, 490, L33

				\reference{}
Faber, S. M., Wegner, G., Burstein, D., Davies, R. L., Dressler, A., Lynden-
Bell, D., 
    Terlevich, R. J., 1989, \apjs, 69, 763

                                \reference{}
Fukazawa, Y., Ohashi, T., Fabian, A. C., Canizares, C. R., Ikebe, Y., Makishima, 
	K., Mushotzky, R. F. \& Yamashita, K. 1994, \pasj, 46, L55

                                \reference{}
Fukazawa, Y., Makishima, K., Tamura, T., Ezawa, H., Xu, H., Ikebe, Y., Kikuchi,
 K. \& Ohashi, T., 1998, \pasj, 50, 187

                                \reference{}
Gibson, B. K., Loewenstein, M. \& Mushotzky, R. F. 1997, \mnras, 290, 623

                                \reference{}
Gunn, J. E. \& Gott, J. R. III 1972, \apj, 176, 1

                                \reference{}
Heckman, T. M. 1981, \apjl, 250, L59
 
                                \reference{}
Ikebe, Y., Makishima, K., Ezawa, H., Fukazawa, Y., Hirayama, M., Honda, H.,
	Ishisaki, Y., Kikuchi, K., Kubo, H., Murakami, T., Ohashi, T. 
	\& Takahashi, T. 1997, \apj, 481, 660

 \reference{}
Irwin, J. \& Bregman, J. N. 1999, submitted to \apj

                                \reference{}
Ishimaru, Y. \& Arimoto, N. 1997, \pasj, 49, 1

                                \reference{}
Kaastra, J. S. 1992, {\it An X-Ray Spectral Code for Optically Thin Plasmas},
    (Internal SRON-Leiden Report, updated version 2.0)

                                \reference{}
Koyama, K., Takano, S. \& Tawara, Y. 1991, \nat, 350, 135
 
                                \reference{}
Kowalski, M. P., Cruddace, R. G., Snyder, W. A., Fritz, G. G., Ulmer, M. P. \& 
	Fenimore, E. E. 1993, \apj, 412, 489

                                \reference{}
Larson, R. B. \& Dinerstein, H. L. 1975, \pasp, 87, 911
 
                                \reference{}
Liedahl, D. A., Osterheld, A. L. \& Goldstein, W. H. 1995, \apjl, 438, L115

                                \reference{}
Loewenstein, M. \& Mathews, W. G. 1991, \apj, 373, 445

				\reference{}
Loewenstein, M., Mushotzky, R. F., Tamura, T., 
     Ikebe, Y., Makishima, K., Matsushita, K., 
      Awaki, H. \& Serlemitsos, P. J. 1994, \apjl, 436, L75

				\reference{}
Loewenstein, M. \& Mushotzky, R. F.  1996, \apj, 466, 695

				\reference{}
Loewenstein, M. \& Mushotzky, R. F  1997, astro-ph/9710339

				\reference{}
Matsumoto, H., Koyama, K., Awaki, H., Tsuru, T., Loewenstein, M., \& Matsushita, K.
, 1997, \apj, 482, 133

				\reference{}
Matsushita, K. 1997, Ph.D. thesis, University of Tokyo

				\reference{}
Matsushita, K., Makishima, K., Awaki, H.;, Canizares, C. R., Fabian, A. C., 
 Fukazawa, Y.,Loewenstein, M., Matsumoto, H., Mihara, T., Mushotzky, R. F., 
 Ohashi, T., Ricker, G. R., Serlemitsos, P. J., Tsuru, T., 
 Tsusaka, Y., \& Yamazaki, T. 1994, \apjl, 436, L41

				\reference{}
Matsushita, K., Makishima, K., Rokutanda, E., Yamazaki, N. Y., \& Ohashi, T., 
1997, \apjl, 488, L125

				\reference{}
Matsushita, K., Makishima, K., Ikebe, Y., Rokutanda, E., 
    Yamazaki, N. Y., \& Ohashi, T., 1998, \apjl, 499, L13

                                \reference{}
Mewe, R., Gronenschild, E. H. B. M. \& Van den Oord, G.  H. J. 1985, \aaps, 62, 
197
 
                                \reference{}
Mewe, R., Lemen, J. R. \& Van den Oord, G. H. J. 1986, \aaps, 65, 511

                                \reference{}
Morrison, R. \& McCammon, D. 1983, \apj, 270, 119

                                \reference{}
Mushotzky, R. F. 1984, \physscr, T7, 157

                                \reference{}
Mushotzky, R. F. 1995, in {\it New Horizons of X-ray Astronomy}, ed. Makino \& 
T. Ohashi 
	(Tokyo: Universal Academy), 243
	  	
                                \reference{}
Mushotzky, R. F. \& Loewenstein, M. 1997, \apjl, 481, L63

                                \reference{}
Mushotzky, R. F., Loewenstein, M., Arnaud, K. A., Tamura, T., Fukazawa, Y., 
	Matsushita, K., Kikuchi, K. \& Hatsukade, I. 1996, \apj, 466, 686

                                \reference{}
Mushotzky, R. F. \& Szymkowiak, A. E. 1988, in {\it Cooling flows in clusters 
and 
	galaxies}, ed. A. C. Fabian (Dordrecht: Kluwer), p. 53-62
 
                                \reference{}
Nagataki, S. \& Sato, K. 1998, \apj, 504, 629

                                \reference{}
Nagataki, S., Hashimoto, M., Sato, K., Yamada, S. 1998, \apj, submitted

                                \reference{}
Nepveu, M. 1981, \aap, 101, 362

                                \reference{}
Nolthenius, R. 1993, \apjs, 85, 1

                                \reference{}
Nomoto, K., Iwamoto, K., Nakasato, N., Thielemann, F.-K., Brachwitz, F.,
    Tsujimoto,  T., Kubo, Y. \& Kishimoto, N., 1997a, Nuclear Physics A, Vol. 
A621

                                \reference{}         
Nomoto, K., Hashimoto, M., Tsujimoto, T., Thielemann, F.-K., Kishimoto, N., \& 
Kubo, 
	Y.  1997b, Nuclear Physics A, Vol. A616

                                \reference{} 
Nomoto, K., Thielemann, F.-K. \& Yokoi, K. 1984, \apj, 286, 644        

                       		\reference{}
Nulsen, P. E. J., Stewart, G. C., Fabian, A. C., Mushotzky, R. F., Holt, S. S., 
Ku, 
	W. H.-M. \& Malin, D. F. 1982, \mnras, 199, 1089


				\reference{}
Pislar, V., Durret, F., Gerbal, D., Lima Neto, G. B. \& Slezak, E. 1997, \aap, 
322, 53
 
                                \reference{}
Ponman, T. J., Bertram, D., Church, M. J., Eyles, C. J., Watt, M. P., Skinner, 
G. K.
	\& Willmore, A. P. 1990, \nat, 347, 450
 
                                \reference{}
Renzini, A., Ciotti, L., D'Ercole, A. \& Pellegrini, S. 1993, \apj, 419, 52
 
                                \reference{}
Schombert, J. M. 1986, \apjs, 60, 603

 
                                \reference{}
Thielemann, F.-K., Nomoto, K. \& Yokoi, K. 1986, \aap, 158, 17
 
                                \reference{}
Thomas, P. A., Fabian, A. C. \& Nulsen, P. E. J. 1987, \mnras, 228, 973

                                \reference{}
Ulmer, M. P., Cruddace, R. G., Fritz, G. G., Snyder, W. A., Fenimore, E. E.
    1987, ApJ, 319, 118

				\reference{}
Valentijn, E. A. 1983, \aap, 118, 123

                                \reference{}
White, R. E. III 1991, \apj, 367, 69

                                \reference{}
White, R. E. III \& Davis, D. S. 1997, in {\it Galactic \& Cluster Cooling 
Flows}, 
     ed. N. Soker, {\it A.S.P. Conf. Series Vol. 115} (San Francisco: A.S.P.) 
pp. 217-226 

				\reference{}
White, R. E. III \& Davis, D. S. 1999, in preparation.

				\reference{}
White, R. E. III, Day, C. S. R., Hatsukade, I., \& Hughes, J. P. 1994, \apj, 
433, 583
 
                                \reference{}
Woosley, S. E. \& Weaver, T. A. 1995, \apjs, 101, 181

                                \reference{}
Xu, H., Ezawa, H., Fukazawa, Y., Kikuchi, K., Makishima, K., Ohashi, T. \& 
	Tamura, T., 1997, \pasj, 49, 9

                                \reference{}
Zabludoff, A., Huchra, J. P.,  \& Geller, M. J. 1990, \apjs, 74, 1
 
				\end{references}
                                
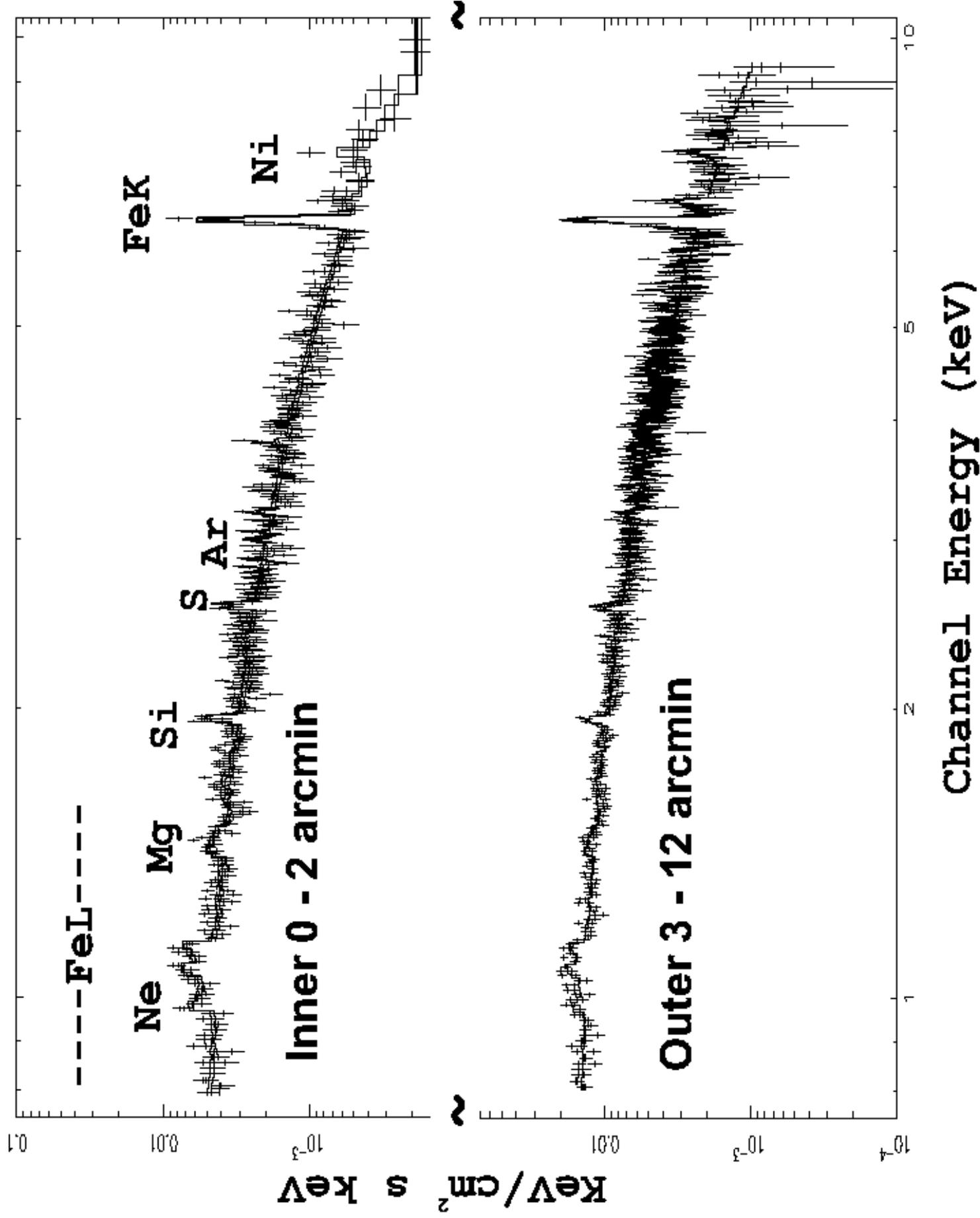
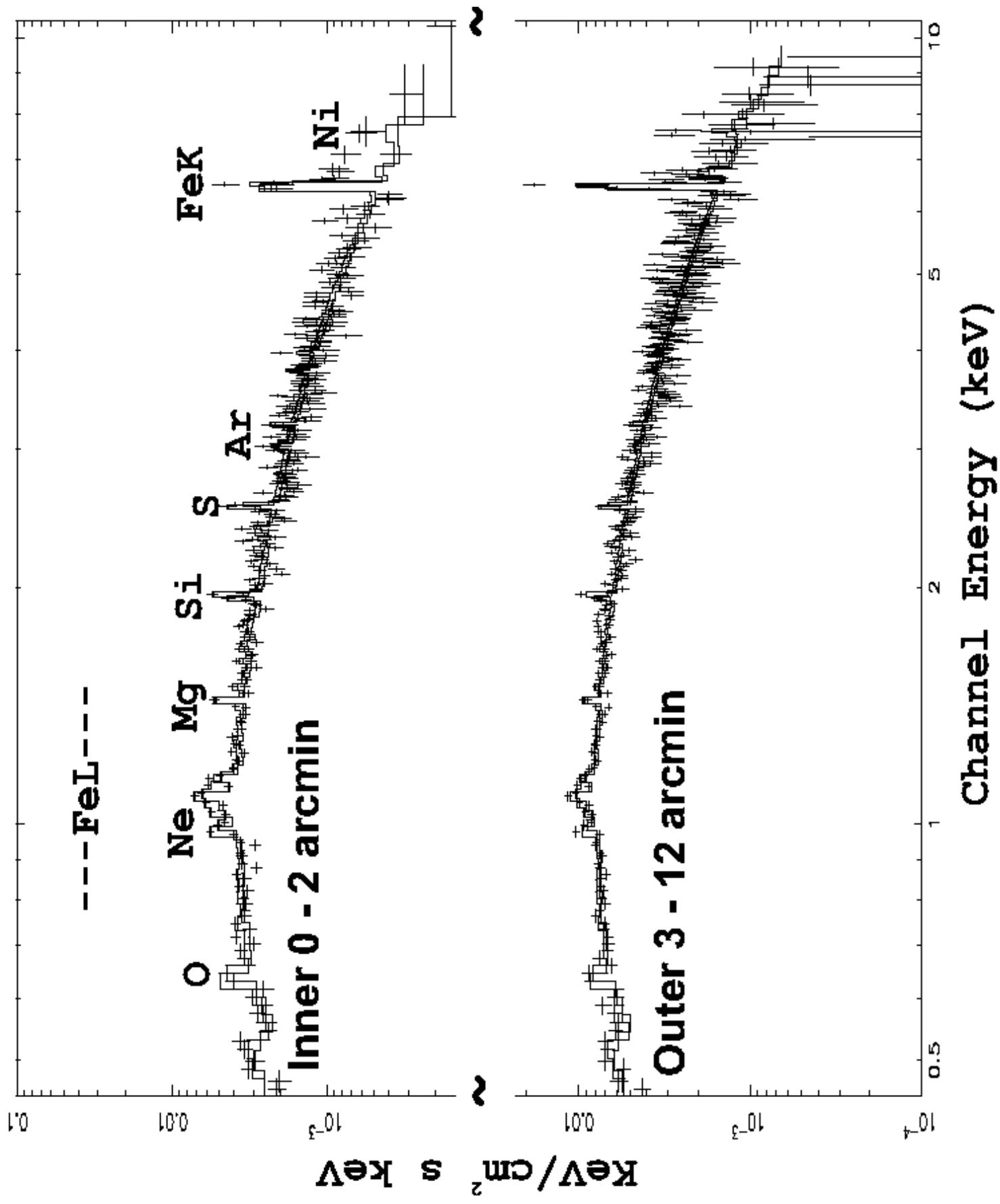
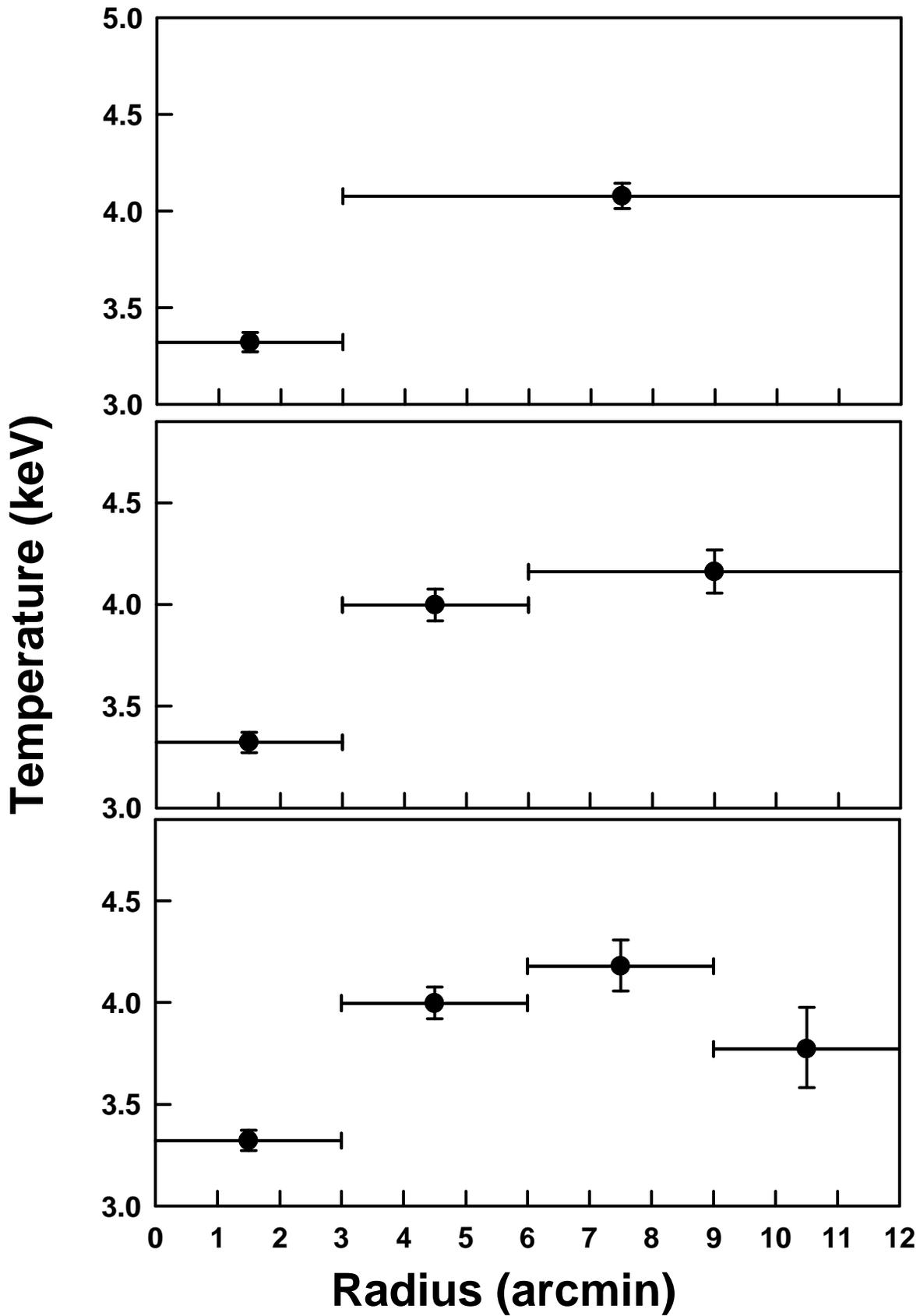
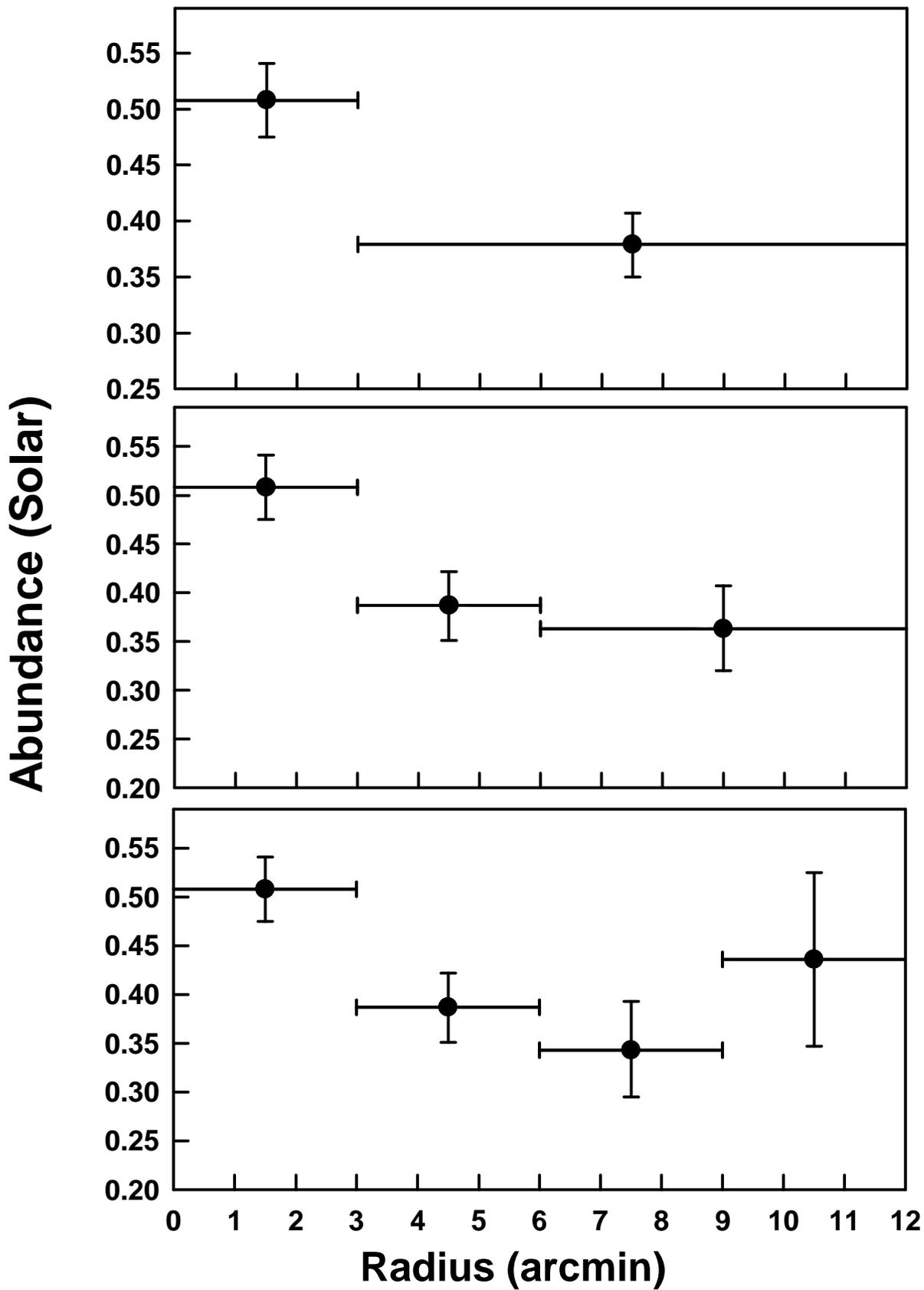
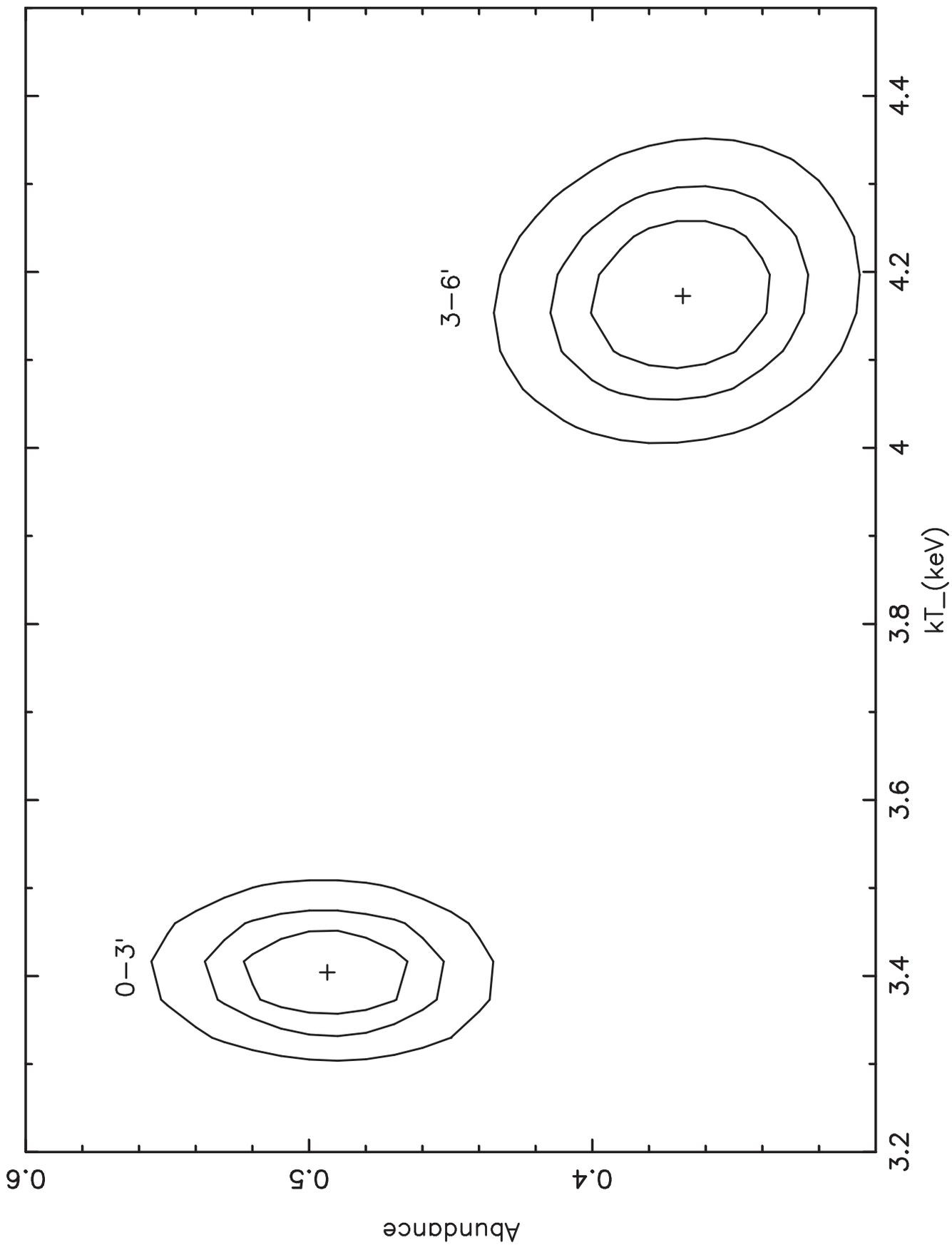
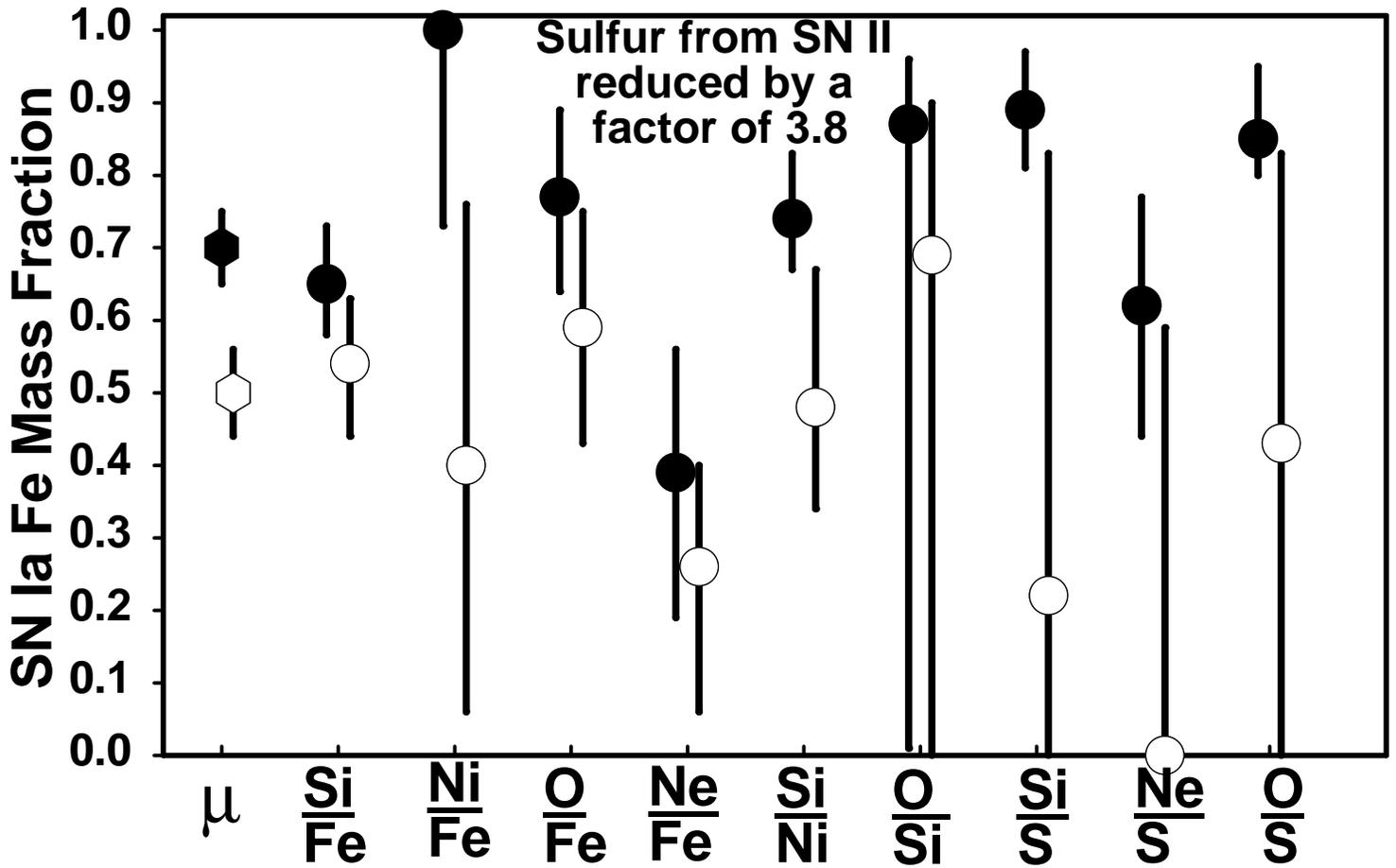
\end{document}